\documentclass[prl, longbibliography, twocolumn, showpacs, preprintnumbers,amsmath,amssymb,superscriptaddress]{revtex4-1}
\usepackage{graphicx}
\usepackage{xcolor}
\usepackage{dcolumn}
\usepackage{bm}

\begin{document}


\title{Linker-mediated self-assembly of mobile DNA-coated colloids}

\author{Xiuyang Xia}
\affiliation{Chemical Engineering, School of Chemical and Biomedical Engineering, Nanyang Technological University, 62 Nanyang Drive, Singapore 637459}
\affiliation{Division of Physics and Applied Physics, School of Physical and Mathematical Sciences,Nanyang Technological University, 21 Nanyang Link, Singapore 637371}%
\author{Hao Hu}
\affiliation{Chemical Engineering, School of Chemical and Biomedical Engineering, Nanyang Technological University, 62 Nanyang Drive, Singapore 637459}
\affiliation{School of Physics and Materials Science, Anhui University, Hefei 230601, People’s Republic of China}
\author{Massimo Pica Ciamarra}
\email{massimo@ntu.edu.sg}
\affiliation{Division of Physics and Applied Physics, School of Physical and Mathematical Sciences,Nanyang Technological University, 21 Nanyang Link, Singapore 637371}
\author{Ran Ni}
\email{r.ni@ntu.edu.sg}
\affiliation{Chemical Engineering, School of Chemical and Biomedical Engineering, Nanyang Technological University, 62 Nanyang Drive, Singapore 637459}

\date{\today}

\begin{abstract}
Developing construction methods of materials tailored for given applications with absolute control over building block placement poses an immense challenge. DNA-coated colloids offer the possibility of realising programmable self-assembly, which, in principle, can assemble almost any structure in equilibrium, but remains challenging experimentally. Here, we propose an innovative system of linker-mediated mobile DNA-coated colloids (mDNACCs), in which mDNACCs are bridged by the free DNA linkers in solution, whose two single-stranded DNA tails can bind with specific single-stranded DNA receptors of complementary sequence coated on colloids. We formulate a mean-field theory efficiently calculating the effective interaction between mDNACCs, where the entropy of DNA linkers plays a nontrivial role. Particularly, when the binding between free DNA linkers in solution and the corresponding receptors on mDNACCs is strong, the linker-mediated colloidal interaction is determined by the linker entropy depending on the linker concentration. 
\end{abstract}

\maketitle

{\section{Teaser}
Temperature-insensitive specific colloidal interactions are found in linker-mediated DNA-coated colloids.}

\section{Introduction}
The ultimate goal of self-assembly is programming many distinct building blocks, and each of them occupies a specific location within a self-assembled structure~\cite{Jacobs:2016kg}.
The recent development of DNA nanotechnology offers possibilities of programmable self-assembly using the specific hybridization between single stranded DNAs (ssDNAs)~\cite{alivisatos1996organization,mirkin1996dna,seemanjacs1996,dnareviewscience}. This works very well in programmable self-assembly of DNA bricks, in which a variety of designed superstructures consisting of thousands of preprogrammed DNA bricks were fabricated~\cite{dnabrickscience,yinpeng2017,Jacobs:2015gi,Sajfutdinow2018}.
However, similar ideas were not well applied in the designed self-assembly of DNA-coated colloids (DNACCs). 
One of the major challenges is that typically each colloid is coated with many DNA linkers, and the effective colloidal interaction mediated by DNA hybridization changes abruptly with temperature, which makes the system difficult reach the equilibrium ordered state~\cite{Jacobs:2016kg}. Thus, only a few groups were able to obtain 3D crystals of DNACCs~\cite{crocker2006,gang2008,mirkin2011,crocker2012,wang2015,wang2015natcomm,gang2015,crocker2017}.

This is particularly detrimental for designed self-assembly of colloidal superstructures, where the temperature window for high-yield self-assembly narrows down quickly with the increasing structure size~\cite{zeravcic2014size}. In this work, we propose a new system of linker-mediated mobile DNA-coated colloids (linker-mediated mDNACCs). Experimentally, one can fabricate mDNACCs by grafting DNA linkers onto the lipid-bilayer coated on the colloids~\cite{vanderMeulen:2013dc,vanderMeulen:2015jb}, which makes the grafted linkers mobile on colloidal surface. 
It was found that the phase diagram of conventional mDNACCs is not qualitatively different from the corresponding immobile DNACCs, where the freezing colloidal density drops to zero quickly with increasing the binding strength between ssDNAs~\cite{Hu:2018ev}. Here  we find that if the ssDNAs grafted on different colloids do not bind with each other directly, but rather through the bridging of free DNA linkers in solution, the effective colloidal attraction does not diverge at the strong binding limit, i.e., low temperature limit. The reason is due to the special entropic effect in the strong binding limit making the linker-mediated attraction between mDNACCs finite and solely depending on the concentration of free DNA linkers in solution, which can be  well controlled over orders of magnitude experimentally. 

\begin{figure*}[ht]
   \centering 
 \includegraphics[width=1.\textwidth]{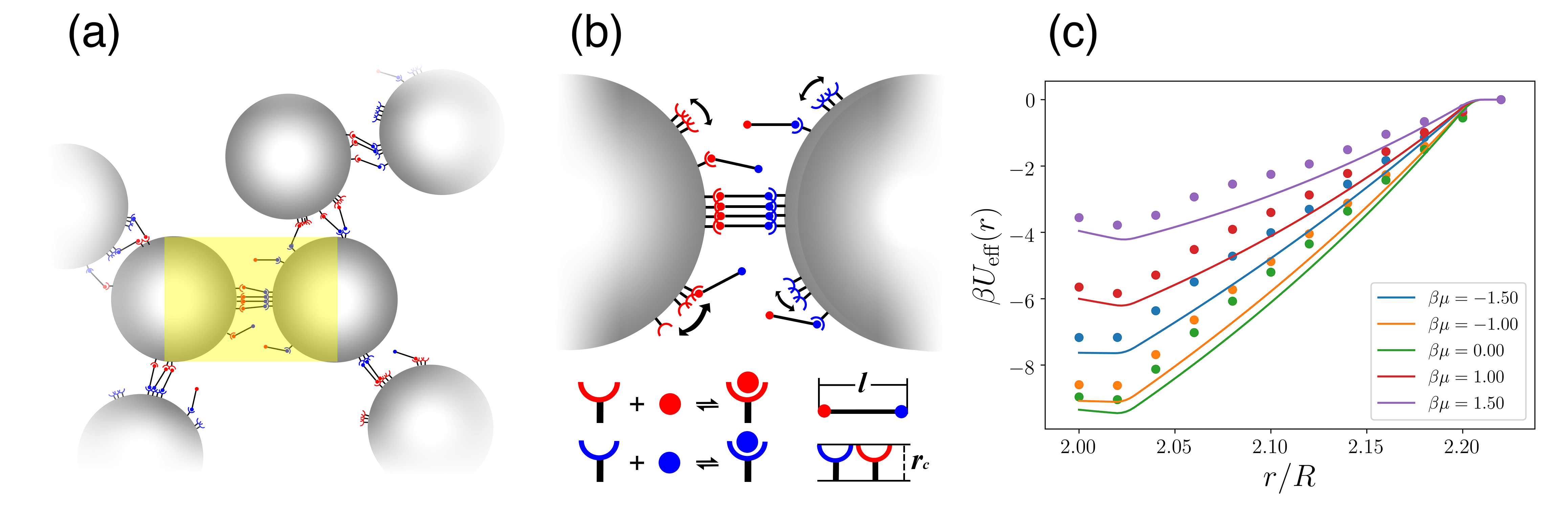}
   \caption{\textbf{Linker-mediated mDNACCs.} (a): Schematic representation of binary mNDACCs, in which the colloidal interaction is mediated via the bridging of free DNA linkers that bind with the mobile ssDNA receptors on mDNACCs; (b): a magnification of the yellow region in (a). Here mobile ssDNA receptors (red or blue) coated on mDNACCs can bind with the ssDNA tails (red or blue spheres) of free linkers of the same color.
(c): The linker-mediated effective interaction $\beta U_{\mathrm{eff}}(r)$ as a function of the center-to-center distance between two mDNACCs  $r$ for different $\beta\mu$ with $\beta\Delta G_{bind}=-3.0$, $n_{i}=50$, $l = 0.2R$, and $r_c=0.01R$. Solid lines are from Eq.~\ref{eff_int}, and symbols are from the direct Monte Carlo (MC) simulations with explicit linkers. }
   \label{fig1}
\end{figure*}

\section{Results} 
\subsection{Model and mean field theory}
We consider an equimolar $AB$-type binary system of volume $V$ consisting of $N$ mDNACCs (hard spheres) of radius $R$ in the solution containing free DNA linkers of chemical potential $\mu$. Each mDNACC $i$ is coated with $n_{i}$ $A$ or $B$ type mobile ssDNA receptors of length $r_c$, which can bind to the ssDNA tail of complementary sequence on a free DNA linker in solution with the binding free energy $\Delta G_{bind}$ (Fig.~\ref{fig1}a). Free DNA linkers are modelled as infinitely thin hard rods of length $l$ with two ssDNA ends of length $r_c$ (Fig.~\ref{fig1}b). Here we assume $R \gg l \gg r_c$, and the free DNA linkers in solution and the mobile ssDNA receptors move much faster than colloids. During the motion of colloids, all linkers and receptors reach equilibrium quickly. One can write down the partition function and calculate the free energy of the linker system using the saddle point approximation. This can be used as the effective interaction between mDNACCs mediated by linkers~\cite{AngiolettiUberti:2014kl,angioletti2013communication,Varilly:2012gl}, which has the contributions from both bonded linkers on mDNACCs and unbound free linkers in solution. 
If we assume that the excluded volume effect between the grafting sites of ssDNA receptors on the same particle can be approximated using the hard-disk like repulsion on a 2D surface, when the area fraction of the grafted ssDNA receptors on the particle is less than $5\%$, the compressibility factor is less than 1.1~\cite{eoshd}, and the interaction between the grafting sites can be approximated as an ideal gas. 
Moreover, we assume that the concentration of DNA linkers in solution is low, and except the binding between complementary ssDNAs, the interaction between them is negligible.  These assumptions are typical in experiments.
Thus, the grand canonical partition function for the bonded DNA linkers on mDNACCs is
\begin{widetext}
\begin{equation}\label{partition_func}
Z(\{m_i,q_{ij}\})=\sum_{\{m_i,q_{ij}\}}
								 W(\{m_i,q_{ij}\})
								\xi_a^{\sum_i m_i}
								\xi_b^{\sum_i \sum_{j>i}q_{ij}}
								e^{\beta\mu (\sum_{i}m_{i}+\sum_{i}\sum_{j>i}q_{ij})},
\end{equation}
\end{widetext}
where $m_{i}$ and $q_{ij}$ are the numbers of linkers bonded with the receptors on particle $i$ with one free end and linkers bridging between particles $i$ and $j$, respectively. Here $\beta = 1/k_B T$ with $k_B$ the Boltzmann constant and $T$ the temperature of the system, respectively, {and $\mu = k_BT \log \rho$ with $\rho$ the concentration of free linkers in the reservoir}. $W(\{m_i,q_{ij}\})$ accounts for all possible combinations of hybridization of $\{m_i,q_{ij}\}$ (see Supplementary Materials S1):
\begin{equation}\label{weq}
W(\{m_i,q_{ij}\})=\prod_i
								\frac{n_i!}{m_i!(n_{i}-m_i-\sum_j q_{ij})!\prod_{j>i}q_{ij}!},
\end{equation}
where $\xi_a$ and $\xi_b$ are the partition functions for the states of linkers only bonded to one mDNACC and bridging between two mDNACCs, respectively. 
Linkers bonded to mDNACCs can stay in two different states: (i) state $a$, (only one end of the linker is bonded to an mDNACC with the other end unbound); (ii) state $b$ (two ends of the linker are bonded to two different mDNACCs).
We introduce a reference linker state $a'$ in the dilute limit of mDNACCs, in which particles are at infinite distance from each other, so that bridges cannot form, and 
the linker is in state $a$ but not interacting with other mDNACCs. Its partition function is $\xi_{a'} = V_{a'}\exp\left(-\beta \Delta G_{bind} \right)$ with $V_{a'}$ the configurational volume that the linker in state $a'$ can explore. At a finite colloidal concentration, the existence of neighbouring colloids influences the free volume of linkers in state $a$, which induces a repulsive free energy $F_{rep}$, and $\xi_a = \xi_{a'} \exp(-\beta F_{rep})$. Similarly, for the bridging linkers in state $b$, $\xi_b = \xi_{a'} \exp[ -\beta (\Delta G_{bind} + F_{cnf})]$, where $F_{cnf}$ is the conformational free energy of the linker bridging between two mDNACCs. Here $F_{rep}$ and $F_{cnf}$ can be calculated exactly at $r_c \rightarrow 0$ for systems of rigid DNA linkers, and otherwise computed with Monte Carlo (MC) simulations for semi-flexible polymeric linkers (see Supplementary Materials S2).

Using the free energy of the bonded linkers $\mathcal{F}(\{m_{i},q_{ij}\})$, we can re-write Eq.~\ref{partition_func} into $Z=\sum_{\{m_i,q_{ij}\}}\exp{(-\beta \mathcal{F}(\{m_{i},q_{ij}\})})$, and with the saddle point approximation $\partial \mathcal{F}(\{m_i,q_{ij} \})/
\partial\{m_i,q_{ij}\}=0$, we obtain
\begin{equation}
\left\{
\begin{aligned}
&m_i = \bar{n}_i\xi_a e^{\beta \mu},\\
&q_{ij}=\bar{n}_i\bar{n}_j \xi_b e^{\beta \mu},
\end{aligned}
\right.
\label{Eq:m_and_x}
\end{equation}
with $\bar{n}_i=n_i-m_i-\sum_i q_{ij}$ the number of unbound free ssDNA receptors on particle $i$. 
{$\bar{p}_{i}$ is the probability to find an unbound ssDNA receptor on particle $i$}, and 
 \begin{equation}
    \bar{p}_{i}+\frac{m_{i}}{n_i}+\sum_{j}\frac{q_{ij}}{n_i}=1.
    \label{Eq:vacancy_balance}
 \end{equation}
Combining Eq.~\ref{Eq:m_and_x} and \ref{Eq:vacancy_balance}, we obtain a set of self-consistent equations:
 \begin{equation}
   \frac{1}{\bar{p}_{i}}=1+\xi_a e^{\beta \mu}+\sum_j \bar{p}_j n_j\xi_b e^{\beta \mu}.
\label{Eq:self_consi}
\end{equation}
Then the free energy of bonded DNA linkers at the saddle point is
\begin{equation}\label{effFF}
\beta F = \sum_{i} \left(n_{i}\log \bar{p}_i + \frac{1}{2}\sum_{j} q_{ij} \right),
\end{equation}
where $\bar{p}_i$ and $q_{ij}$ are the solution to Eq.~\ref{Eq:self_consi} (see Supplementary Materials S3). 
{This resulting effective potential shares the same form as the one in conventional mDNACCs~\cite{AngiolettiUberti:2014kl}, while the physics is different.}

\begin{figure*}[ht]
   \centering
   \includegraphics[width=1.\textwidth]{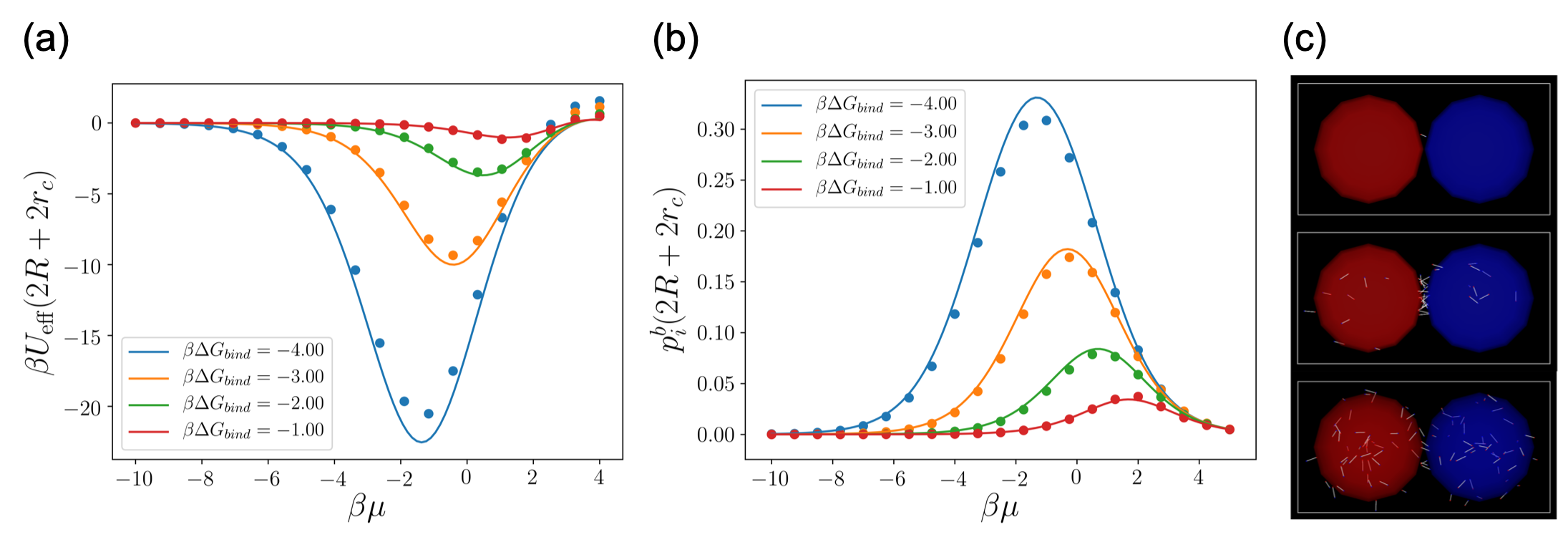}
   \caption{\textbf{Re-entrant melting of linker-mediated mDNACCs.} (a,b): $\beta U_{\mathrm{eff}}(2R+2r_c)$ (a) and $p_{i}^b(2R+2r_c)$ (b) as a function of $\beta \mu$ between two mDNACCs with $n_{i} = 50$ for various $\beta\Delta G_{bind}$. The solid lines are from theoretical prediction (Eq.~\ref{eff_int}), and the symbols are from direct simulations with explicit DNA linkers. (c): typical snapshot for direct simulations of two linker-mediated mDNACCs with explicit DNA linkers with $\beta\Delta G_{bind}=-4.0$ at $\beta \mu = -8.05, -1.90$ and 2.0 (from top to bottom), where the (red and blue) big spheres and white line segments are mDNACCs and DNA linkers, respectively.}
   \label{fig2}
\end{figure*}

Besides the free energy contribution from the linkers bonded on mDNACCs, the unbound free linkers in solution also contributes a depletion effect $U_{dep}$ ~\cite{depletion_AO,depletionbook} (see Supplementary Materials S2), and the resulting linker-mediated effective interaction between mDNACCs can be written as 
\begin{equation}\label{eff_int}
\beta U_{\mathrm{eff}} = \sum_{i} \left[n_{i}\log\left(\frac{\bar{p}_i}{\bar{p}_i'} \right) + \frac{1}{2}\sum_{j} q_{ij} \right] + \beta U_{dep},
\end{equation}
where $\bar{p}_i'$ is the probability of finding an unbound ssDNA receptor on an isolated particle $i$, i.e., no bridge formed on particle $i$, in the reservoir of free linkers with chemical potential $\mu$. {We note that the mean field approach employed here is only meaningful if $\Delta G_{bind}$ is on the scale of a few $k_BT$~\cite{AngiolettiUberti:2014kl,angioletti2013communication,Varilly:2012gl}, otherwise kinetic effects need to be taken into account~\cite{kinetic2016a,kinetic2016b}. To check the validity of our mean field results in experiments, one can compare the time it takes for a single colloid to diffuse its own radius $t_{diffusion}$ with the time scale of bond formation $t_{on}$ and bond breaking $t_{off}$, and ensure $t_{diffusion} \gg t_{on} + t_{off}$.} 

\begin{figure}[ht]
   \centering
   \includegraphics[width=.45\textwidth]{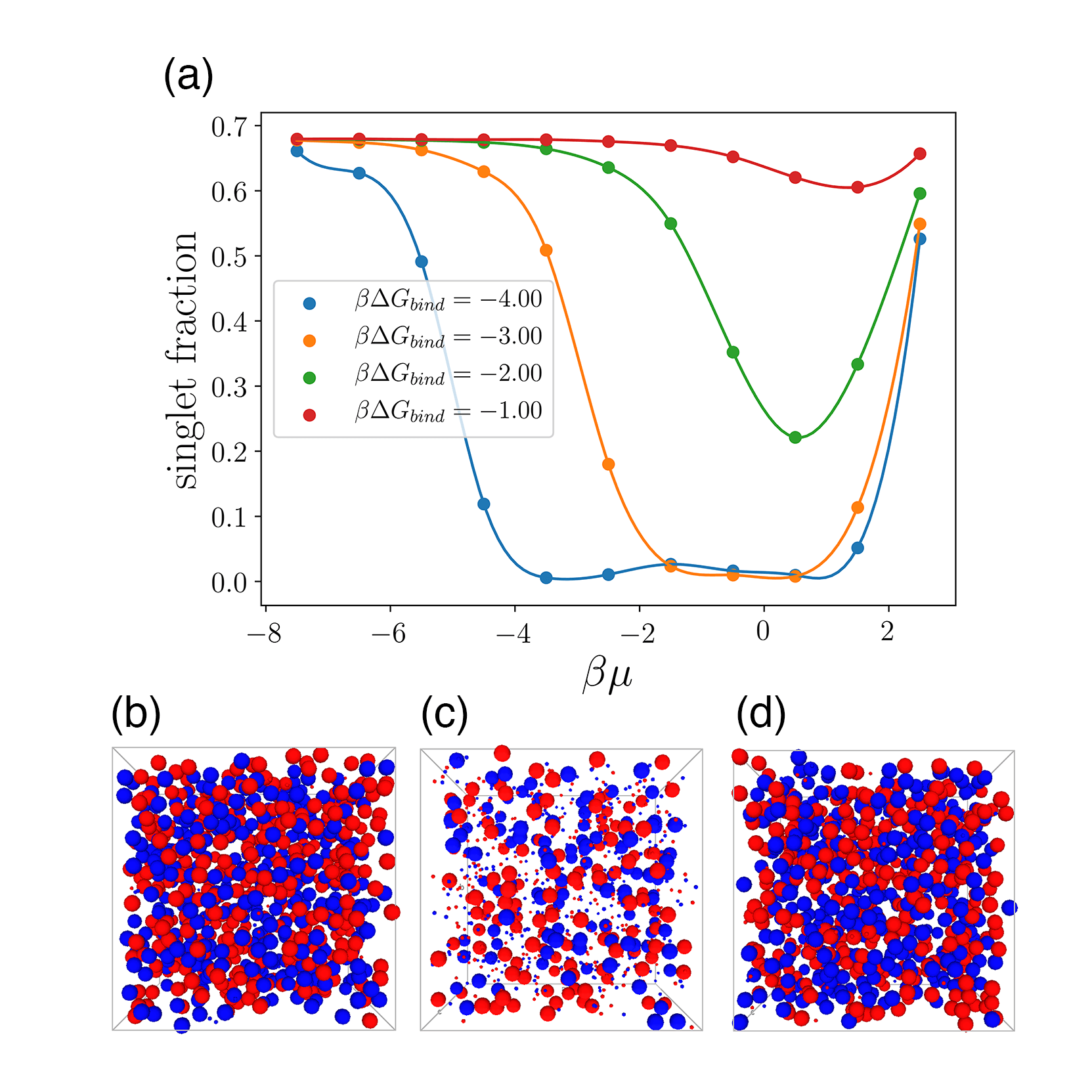}
   \caption{\textbf{Simulation of re-entrant melting.} (a): Singlet fraction in binary linker-mediated mDNACCs (packing fraction $\eta=0.1$) as a function of $\beta\mu$ for different $\beta\Delta G_{bind}$ (with lines as a guide to the eye). (b-d): The typical snapshots of the system at $\beta \mu = -6.5$ (b), $-2.5$ (c), and 2.5 (d), where non-singlets are drawn ten times smaller. Here $n_{i}=50$ and $\beta\Delta G_{bind}=-4.0$ at the packing fraction $\eta = 0.1$.}
   \label{fig3}
\end{figure}

\subsection{Numerical verification of the mean field theory}
We perform MC simulations with explicit DNA linkers to verify Eq.~\ref{eff_int}. 
As the system spans multiple length scales, direct simulations of many mDNACCs with explicit DNA linkers are prohibitively expensive, and we choose to perform thermodynamic integrations (see Methods) to calculate the linker-mediated effective interaction between two different mDNACCs (like in Fig.~\ref{fig2}c). We plot the calculated linker-mediated effective interaction $\beta U_{\mathrm{eff}} (r)$ between two colloids coated with the same number but different types of mobile ssDNA receptors compared with the theoretical prediction of Eq.~\ref{eff_int} in Fig.~\ref{fig1}c for various $\mu$. One can see that the numerically calculated effective interaction between two mDNACCs agrees well with the theoretical prediction, and Eq.~\ref{eff_int} generally predicts a slightly stronger attraction than the one measured in direct simulations. 
{The reason causing the discrepancy is two-fold. First, it is known that the mean field approach developed here generally predicts a slightly stronger attraction between mDNACCs compared with direct simulations with explicit linkers, as the mean field theory assumes that the probability for any ssDNA receptor coated on colloids to be unbound is uncorrelated to the probability of a ssDNA tail on a free linker to be unbound~\cite{titojcp,Varilly:2012gl}. The difference is visible when the number of ssDNA receptors or the free DNA linkers in the system is very small, which is not the case in our simulations.} 
Second, the analytical form of $F_{cnf}$ is only exact at $r_c \rightarrow 0$, while for finite small $r_c$, it slightly overestimates the conformational entropy for bridging linkers. 
{We believe that the overestimation of $F_{cnf}$ is the major (but not the only) reason responsible for the overestimation on the amount of bridging linkers and stronger attraction (see Supplementary Materials S2).} 

As shown in Fig.~\ref{fig1}c, $\beta U_{\mathrm{eff}} (r)$ is negative and attractive at $2R < r < 2R + l$ with the minimum located around $2R+2r_c$, whose magnitude indicates the strength of attraction between two mDNACCs. Interestingly, as shown in Fig.~\ref{fig2}a, with increasing $\beta \mu$ from $-10$ to about 0, the attraction between two mDNACCs first becomes stronger, i.e., $\beta U_{\mathrm{eff}} (2R+2r_c)$ becomes more negative, while further increasing $\beta \mu$ makes the attraction weaker. Simultaneously, the probability of forming bridges between two colloids $p_{i}^{b}(2R+2r_c) = q_{ij}/n_{i}$ first increases and then decreases with increasing $\mu$ (Fig.~\ref{fig2}b). Typical snapshots from direct simulations at various $\mu$ (Fig.~\ref{fig2}c) show that at both very low and high linker concentrations, there are very few bridges formed between colloids, while many bridges form at certain intermediate linker concentration. It is easy to understand that very few bridges form at small $\mu$, as there are limited linkers available to bridge between mDNACCs, while it is not trivial that the number of bridges decreases at large $\mu$. To understand this, we refer to Eq.~\ref{Eq:m_and_x}, which implies 
\begin{equation}\label{reentrant}
\frac{\sum_j q_{ij}}{m_i}=\frac{\sum_j \bar{n}_j \xi_b}{\xi_a}.
\end{equation}
When $\mu \rightarrow \infty$, all ssDNA receptors on mDNACCs are bonded, i.e., $\bar{n}_j \rightarrow 0$, and $\xi_a$ and $\xi_b$ are finite numbers and do not depend on $\mu$, which implies ${\sum_j \bar{n}_j \xi_b}/{\xi_a} \rightarrow 0$. On the left side of Eq.~\ref{reentrant}, $m_{i}$ changes with $\mu$ but remains finite leading to $\sum_j q_{ij} \rightarrow 0$, which suggests that at very high linker concentration, there is no bridge formed between mDNACCs and explains the drop of $p_{i}^{b}(2R+2r_c)$ at large $\mu$. 
{The physical explanation is that at $\mu \rightarrow \infty$, mDNACCs tend to absorb as many free linkers as possible to minimize the $-\mu (\sum_i m_i + \sum_i \sum_{j>i} q_{ij})$ term in the grand potential of the system by maximizing $\sum_i m_i + \sum_i \sum_{j>i} q_{ij}$, which drives every ssDNA receptor to bind with a free linker with no bridge formed between ssDNA receptors, i.e., $q_{ij} = 0$.} 
This suggests a re-entrant melting of mDNACCs with increasing $\mu$. To demonstrate this, we perform MC simulations with the effective interaction of Eq.~\ref{eff_int} for an equimolar binary mixture containing $N=864$ mDNACCs at the packing fraction $\eta  = 4N \pi R^3/3V = 0.1 $ with various $\mu$, where we regard a colloid as a singlet if there is no bridge formed on it. As shown in Fig.~\ref{fig3}, one can see that with increasing $\mu$, the singlet fraction first drops then increases at large $\mu$. 
{It is known that in systems of mDNACCs, multi-body effects are important~\cite{AngiolettiUberti:2014kl}, while the predicted re-entrant melting transition based on the effective pair potential between mDNACCs indeed exists in multi-particle systems. The reason is that Eq.~\ref{reentrant} actually accounts for multi-particle systems, which already includes multi-body effects.}
Similar re-entrant melting was also found in a recent work of linker-mediated immobile DNACCs, {in which the ssDNA receptors are grafted on the colloids instead of being mobile on the particle surface~\cite{lowensohn2019linker}. A local chemical equilibrium approach was employed in Ref~\citep{lowensohn2019linker} to treat the linker-mediated interaction between colloids, and it is related to the statistical mechanical description used here~\cite{Varilly:2012gl}. Moreover, to account for the immobility of the grafting points of ssDNA receptors on colloids, a Derjaguin like approach~\cite{Derjaguin1934} adapted to the presence of free linkers in solution was used in Ref~\citep{lowensohn2019linker}, while as mentioned above, the essential physics driving the re-entrant melting of DNACCs remains the same and does not depend on the mobility of ssDNA receptors on colloidal surface.}

\begin{figure*}[ht]
   \centering
   \includegraphics[width=1.\textwidth]{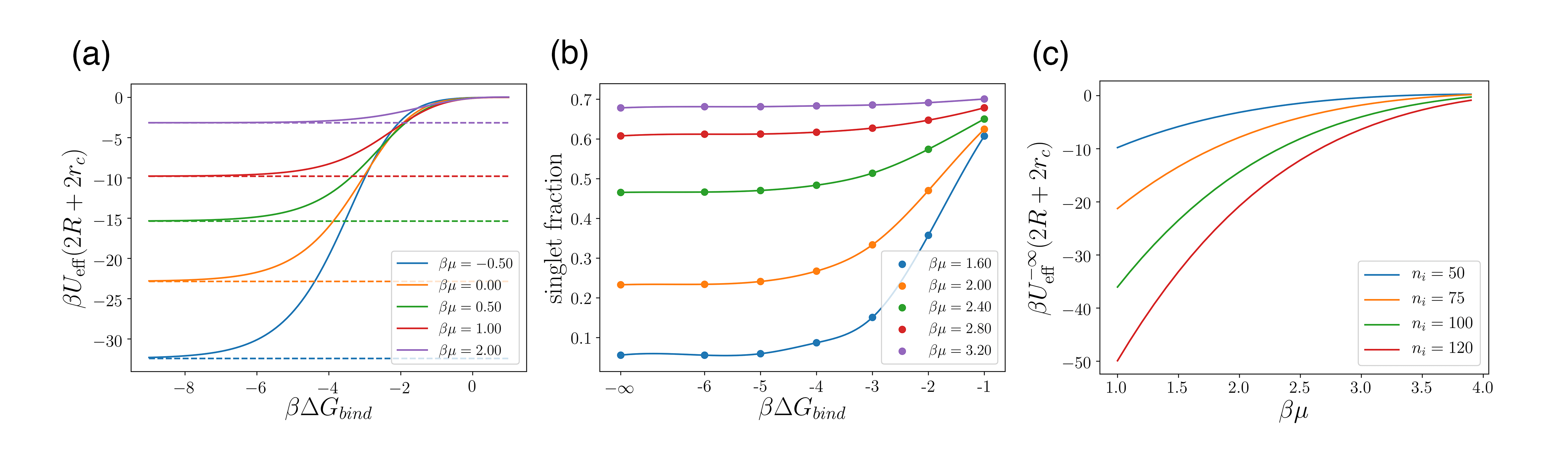}
   \caption{\textbf{Entropy driven linker-mediated mDNACCs at the strong binding limit.} (a): Effective pair interaction $\beta U_{\mathrm{eff}}(2R+2r_c)$ as a function of $\beta\Delta G_{bind}$ for various $\beta \mu$ predicted by Eq.~\ref{eff_int} (solid lines) and ~\ref{eff_inf} (dashed lines).
  (b): Singlet fraction in binary linker-mediated mDNACCs at $\eta=0.1$ as a function of $\beta\Delta G_{bind}$ for different $\beta\mu$ (with lines as a guide to the eye). Here $n_i=50$.
  (c): Effective pair interaction $\beta U_{\mathrm{eff}}^{-\infty}(2R+2r_c)$ as a function of $\beta\mu$ for various $n_i$ at $\beta\Delta G_{bind} \to -\infty$ (Eq.~\ref{eff_inf}).
 }
   \label{fig4}
\end{figure*}

\subsection{Entropic effects in the strong binding limit}
Furthermore, we investigate the effective interaction between mDNACCs with changing the binding strength between DNA linkers and corresponding receptors, i.e., decreasing $\Delta G_{bind}$. One might think that the stronger binding between DNA linkers and receptors would naturally make the attraction between mDNACCs stronger and eventually diverge at $\Delta G_{bind} \rightarrow -\infty$. Intriguingly, however, as shown in Fig.~\ref{fig4}a, at $\Delta G_{bind} \rightarrow -\infty$, $\beta U_{\mathrm{eff}} (2R+2r_c)$ does not diverge but reaches a plateau depending on $\mu$. The MC simulations for systems of many mDNACCs also show that the singlet fraction drops with decreasing $\Delta G_{bind}$ and reach a plateau at the strong binding limit (Fig.~\ref{fig4}b).
To understand this counter-intuitive phenomenon, we modify our mean field theory as follows (see Supplementary Materials S4). When $\Delta G_{bind} \rightarrow -\infty$, all receptors on mDNACCs are occupied, i.e., $\bar{n}_i = 0$, and using the saddle point approximation, one can write down the free energy of the bonded DNA linkers as
\begin{equation}
\begin{aligned}
\beta F_{\inf} & = & \sum_{i} \left[ n_{i} \log \left(\frac{m_{i}}{n_{i} \xi_a e^{\beta (\Delta G_{bind} + \mu)}} \right) \right.\\
& &  \left.+ \frac{1}{2} \sum_{j} q_{ij} + n_{i} \beta \Delta G_{bind} \right],
\end{aligned}
\end{equation}
where
\begin{equation}\label{cond_inf}
q_{ij} = \frac{m_{i} m_{j} \xi_b }{\xi_{a}^2 e^{\beta \mu}} = \frac{m_i m_j}{V_{a'}} e^{\beta \left(2F_{rep} - F_{cnf} - \mu \right)}.
\end{equation}
Here the `enthalpy' term $\sum_{i} n_{i} \beta \Delta G_{bind}$ does not depend on the colloidal configuration and can be neglected, and $\xi_a e^{\beta (\Delta G_{bind} + \mu)} = V_{a'}e^{\beta (\mu - F_{rep})}$ is the entropy part of the partition function, and does not depend on $\Delta G_{bind}$. Therefore, the resulting effective colloidal interaction at the strong binding limit is
\begin{equation}
\begin{aligned}\label{eff_inf}
\beta U_{\mathrm{eff}}^{-\infty} & = & \sum_{i} \left[ n_{i} \log \left(\frac{m_{i}}{n_{i}  V_{a'}e^{\beta (\mu - F_{rep})}} \right) \right.\\
& &  \left.+ \frac{1}{2} \sum_{j} q_{ij} \right] + \beta U_{dep},
\end{aligned}
\end{equation}
which solely depends on entropy.
We plot the prediction of Eq.~\ref{eff_inf} as dashed lines in Fig.~\ref{fig4}a, which quantitatively agree with the converged plateau of $\beta U_{\mathrm{eff}}$ at the strong binding limit. Moreover, in Fig.~\ref{fig4}c, we plot $\beta U_{\mathrm{eff}}^{-\infty}$ as a function of $\mu$ for various $n_{i}$, and one can see that $\beta U_{\mathrm{eff}}^{-\infty}$ increases with increasing $\mu$ or decreasing $n_{i}$. To explain this, we consider the effective pair interaction between two fixed mDNACCs $i$ and $j$ with $n_{i} = n_{j}$ (like in Fig.~\ref{fig2}c) at $\Delta G_{bind} \rightarrow -\infty$, where $m_{i} = m_{j}$, and $m_{i} + q_{ij} = n_{i}$.  Eq.~\ref{cond_inf} implies $q_{ij}/m_{i}^2 = \xi_b / [\xi_{a}^2 \exp(\beta \mu)]$, and $\xi_{b/a}$ does not depend on $\mu$ or $n_{i}$.
With increasing $\mu$, $\xi_b / [\xi_{a}^2 \exp(\beta \mu)]$ decreases, and at fixed $n_{i}$, $q_{ij}$ becomes smaller, which implies less bridges formed between $i$ and $j$, and the less negative $\beta U_{\mathrm{eff}}^{-\infty}$. Similarly, at fixed $\mu$, $q_{ij}/m_{i}^2=\xi_b / [\xi_{a}^2 \exp(\beta \mu)]$ is a constant, and the smaller $n_{i}$ leads to the smaller $q_{ij}$ and the less negative $\beta U_{\mathrm{eff}}^{-\infty}$.

\begin{figure}[ht]
   \centering
   \includegraphics[width=.45\textwidth]{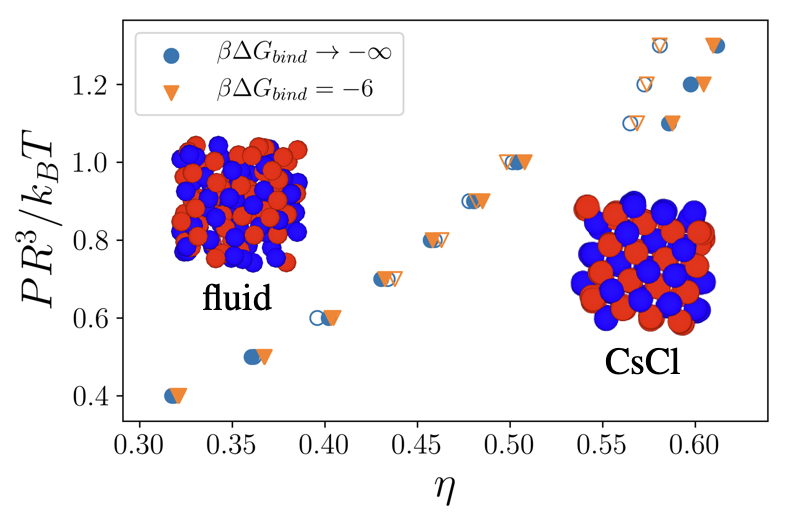}
   \caption{{\textbf{Crystallization of linker-mediated mDNACCs.} Equation of state of the linker-mediated mDNACC systems at $\beta \mu = 1.6$ with different $\beta \Delta G_{bind}$, where the filled and open symbols are the results from $NPT$ simulations expanding from perfect CsCl crystals and compressing from random fluids, respectively. The insets are the snapshots of fluid and nucleated CsCl crystal, respectively.}}
   \label{fig5}
\end{figure}

{It is known that for systems of short range attractive particles, at the infinitely strong attraction limit, vibration entropy stabilizes the floppy crystals~\cite{Hu:2018ev}, which first nucleates from the dilute fluid. Using Eq.~\ref{eff_int} and ~\ref{eff_inf}, we perform isothermal-isobaric ($NPT$) MC simulations for the binary system of linker-mediated mDNACCs at $\beta \mu = 1.6$ at the strong binding, i.e., $\beta \Delta G_{bind} = -6$, and the calculated equation of state (EOS) is shown in Fig.~\ref{fig5}. Compared with the EOS at $\beta \Delta G_{bind} \rightarrow -\infty$, one can see that the random fluid first nucleates into CsCl at the pressure $P R^3/k_BT \simeq 1.1$, and this is qualitatively different from the situation of conventional mDNACC systems, in which the crystallization pressure of the system approaches zero at the strong binding limit~\cite{Hu:2018ev}.}

{\subsection{Generalization to multicomponent systems}}
{In this section, we generalize our mean field theory to calculate the effective interaction in a multicomponent system consisting of $N_c$ types of mDNACCs, in which the total number of mDNACCs is $N$. 
mDNACC $i$ is coated with only one type of ssDNA receptors, e.g., type $I$, and we regard the mDNACCs coated with type $I$ ssDNA receptors as the type $I$ particles. The number of ssDNA receptors coated on mDNACC $i$ is $n_i$.
There are multiple types of free linkers in solution, which are described by a $N_c \times N_c$ connectivity matrix $\mathbf{M}$, and if the free linkers connecting mDNACCs of type $I$ and $J$ exist in the solution, $\mathbf{M}_{IJ} = 1$, otherwise $\mathbf{M}_{IJ} = 0$. We assume that the lengths of all free linkers in solution and all ssDNA receptors coated on mDNACCs are $l$ and $r_c$, respectively. Then the grand canonical partition function of the bonded linker is}
\begin{widetext}
{\begin{equation}\label{general1}
Z(\{m_i,q_{ij}\})=\sum_{\{m_i,q_{ij}\}}
W\left(\{m_i , q_{ij}\}\right)
\prod_I{
\prod_{i}^{i \in N_I}
{\xi_{a,I}}^{\sum_i m_i} e^{\beta\mu_I \sum_i m_i}}
\prod_{I,J}^{\mathbf{M}_{IJ} = 1}
\prod_{i}^{i \in N_I} \prod_{j>i}^{j \in N_J}
 \xi_{b,IJ}^{\sum_{ij} q_{ij}} 
e^{\beta\mu_{IJ}{\sum_{ij} q_{ij}}},
\end{equation}}
\end{widetext}
{where $m_i$ and $q_{ij}$ are the number of linkers bonded on particle $i$ with a free end and the number of bridges formed between particles $i$ and $j$, respectively. $N_I$ consists of all the particles of type $I$ in the system.
$\mu_{IJ}$ is the chemical potential of the linkers that can bridge particles of type $I$ and $J$, and $\mu_I = \sum_{J}^{\mathbf{M}_{IJ} = 1} \mu_{IJ}$. $\xi_{a,I}$ and $\xi_{b,IJ}$ are the partition functions of linkers in the states of bonded on particle type $I$ with a free end (state $a$), and bridging between particles of type $I$ and $J$ (state $b$), respectively, with $\xi_{a,I} = V_{a'} \exp[-\beta (\Delta G_{bind,I} + F_{rep})]$ and $\xi_{a,I} = V_{a'} \exp[-\beta (\Delta G_{bind,I} + \Delta G_{bind,J}+ F_{cnf})]$, where $\Delta G_{bind,I}$ is the binding free energy between a ssDNA receptor of type $I$ coated on mDNACC and another ssDNA tail of complementary sequence on a free linker in solution. $W\left(\{m_i , q_{ij}\}\right)$ accounts for all possible combinations of hybridization of $\{m_i , q_{ij}\}$, and the form is the same as in Eq.~\ref{weq}.
Using the free energy of the bonded linkers $\mathcal{F}\left(\{m_i , q_{ij}\}\right)$, we can re-write Eq.~\ref{general1} into $Z=\sum_{\{m_i,q_{ij}\}} \exp\left( -\beta\mathcal{F}\left(\{m_i, q_{ij}\}\right) \right)$, and with the saddle point approximation $\partial\mathcal{F}\left(\{m_i , q_{ij}\}\right)/\partial\{m_i , q_{ij}\}=0$, we obtain
\begin{equation}
\left\{
\begin{aligned}
m_i=&\bar{n}_i \xi_{a,I}e^{\beta\mu_I}\\
q_{ij}=&\bar{n}_i \bar{n}_j \xi_{b,IJ} e^{\beta\mu_{IJ}}.
\end{aligned}
\right.
\end{equation}
With these, we further obtain a set of self-consistent equations
\begin{equation}\label{effEQ}
\bar{p}_i=\frac{1}
{
\xi_{a,I}e^{\beta\mu_I}+
\sum^{\mathbf{M}_{IJ}=1}_{J}\sum^{j\in N_J}_j
 n_j \bar{p}_j \xi_{b,IJ}e^{\beta\mu_{IJ}}+1
}
\end{equation}
where $\bar{p}_i$ is the probability to find an unbound ssDNA receptor on particle $i$.
Then the free energy of bonded DNA linkers at the saddle point is
\begin{equation}\label{generalF}
\beta F= \sum_{i} \left(
n_i\log{\bar{p}_i}+
\frac{1}{2} \sum_{j} q_{ij}
 \right),
\end{equation}
where $\bar{p}_i$ and $q_{ij}$ are the solution to Eq.~\ref{effEQ}. One can see that Eq.~\ref{generalF} shares the same form as Eq.~\ref{effFF}, and using the same way as proving the positive definiteness of Eq.~\ref{Eq:self_consi} (Supplementary Materials S3), one can prove that Eq.~\ref{effEQ} is also positive definite.}

{Similarly, when $\beta \Delta G_{bind,I} \rightarrow -\infty, 1 \leq I \leq N_c$, one can obtain the free energy of the bonded DNA linkers as
\begin{equation}
\begin{aligned}
\beta F_{\mathrm{inf}} & = &
\sum_{i}\left[ n_{i}
 \log \left(\frac{m_{i}}{n_{i} V_{a'} e^{\beta(\mu_I - F{rep})} }\right)\right.\\
 & & \left. +\frac{1}{2} \sum_{j} q_{i j} +n_i \beta\Delta G_{bind,I} \right],
\end{aligned}
\end{equation}
with
\begin{equation}
q_{ij}  =m_i m_j \frac{e^{\beta(2F_{rep}-F_{cnf})}}{V_{a'}}
e^{- \beta \left(\mu_I + \mu_J - \mu_{IJ} \right)},
\end{equation}
where the `enthalpy' term $\sum_i n_i \beta \Delta G_{bind,I}$ does not depend on the colloidal configuration, and can be neglected. Therefore, in the multicomponent system of linker-mediated mDNACCs, the effective colloidal interaction at the infinitely strong binding limit is not diverging and  solely depends on the entropy of linkers.}

\section{Discussion}
In conclusion, we have proposed a linker-mediated mDNACC system, in which the interaction between mDNACCs is bridged by the free DNA linkers in solution. We formulate a mean field theory to calculate the effective interaction between mDNACCs, which well agrees with numerical simulations with explicit DNA linkers. The mean field theory can be further used to construct MC simulations for efficiently simulating the collective self-assembly of linker-mediated mDNACCs. Furthermore, combining analytic theories with numerical simulations, we find previously unknown entropic effects in linker-mediated mDNACC systems. With increasing the concentration of free DNA linkers from zero, the linker-mediated effective interaction between mDNACCs changes non-monotonically, and the strongest colloidal interaction appears at some intermediate free linker concentration, which induces a re-entrant melting in linker-mediated mDNACCs. Moreover, at fixed free linker concentration, with increasing the binding strength between free linkers in solution and the receptors on mDNACCs, we find that the linker-mediated  attraction between mDNACCs becomes stronger and reaches a plateau at the strong binding limit, i.e., $\Delta G_{bind} \rightarrow -\infty$.
This is due to the fact that at the strong binding limit, all receptors on mDNACCs are bonded. Therefore, whether forming bridges between mDNACCs does not change the `enthalpy' of the system, and the linker-mediated interaction between mDNACCs is dominated by the entropy of DNA linkers, which depends on the concentration of free linkers in solution. 
Here we model the DNA linkers as stiff rods, and to model realistic DNA linkers that are essentially semi-flexible polymers, one just needs to numerically calculate the partition functions $\xi_a$ and $\xi_b$, while the theoretical framework and physics remain the same.
Moreover, we anticipate that the mobility of the grafted ssDNA receptors does not change the qualitative physics. The entropy driven effective colloidal interaction at the strong binding limit should also exist in the recent experimental system of linker-mediated immobile DNA-coated colloids, in which the re-entrant melting transition with increasing the concentration of free DNA linkers was already observed~\cite{lowensohn2019linker}.
As the concentration of free DNA linkers can be well controlled in experiments, this suggests a new way to precisely tune the colloidal interaction for addressable assembly of DNA-coated colloids to avoid the abrupt change of colloidal interaction with temperature in conventional systems of DNACCs. 
{Similar types of entropy driven colloidal interaction have also been observed in other systems of DNACCs, in which the effective interaction between DNACCs is induced by the direct hybridization of ssDNAs coated on colloids, and the general feature of such non-diverging colloidal interactions at the low temperature limit is that when the binding strength between ssDNAs is infinitely strong, the colloidal interaction is a result of the competition between different types of entropies, while the enthalpy does not change with the colloidal configuration~\cite{AngiolettiUberti:2012gi,rogers2015programming}.}
Compared with these existing entropy driven methods for addressable assembly of DNACCs~\cite{rogers2015programming,AngiolettiUberti:2012gi}, the advantage of using linker-mediated mDNACCs is that 
to encode all possible specific interactions between $N$ different mDNACCs would need only $N$ distinct grafted sequences~\cite{lowensohn2019linker} instead of $N(N-1)/2$ in the other systems, which is particularly important for large $N$~\cite{chaikin2012}, and each specific interaction  can be easily switched on or off \emph {in situ} by introducing or {(using strong absorbers) to remove} the corresponding free DNA linkers, {which offers an extra degree of freedom in controlling the pathway of designed assembly.} Moreover, the mobile feature of the receptors coated on colloids makes it possible to derive the closed form for effective interactions between colloids (like Eq.~\ref{eff_int} and \ref{eff_inf}) to efficiently simulate, investigate and design the collective self-assembly, for which direct simulations with explicit linkers are prohibitively expensive.

\section{Methods}
\subsection{Numerical calculation of the effective interaction between two mDNACCs with explicit DNA linkers}
In the simulation box, we place two mDNACCs (hard spheres of radius $R$), and each of them is coated with $n_i$ different types (A or B) of ssDNA receptors of length $r_c$, which can specifically bind with one end of the free DNA linkers of complementary sequences with length $l$.
We consider there is an absorption layer with thickness $r_c$ on particles and the linker ends of complementary sequence in the layer can be bonded with a certain probability. Note that the linker ends can move in the absorption layer freely without any energy cost.
The infinitely thin stiff linkers cannot overlap with the particles, and periodic boundary conditions are applied in all directions.
As typically there are many linkers bonded onto each mDNACC, most of the trial moves of mDNACCs would be rejected. Therefore, in each of our MC simulations, we fix the position of two mDNACCs in the system with the center-to-center distance $r$, and the free energy of the system can be calculated through the thermodynamic integration
\begin{equation}
\begin{aligned}
F(\mu,r) & =   F(-\infty,r) + \int_{-\infty}^{\mu} \left(\frac{\partial F}{\partial \mu'} \right) d \mu' \\
& =   F(-\infty,r) - \int_{-\infty}^{\mu} \langle N_{l}\rangle_{\mu' V T}^r d \mu',
\end{aligned}
\end{equation}
where $\mu$ is the chemical potential of linkers, and $\langle N_l \rangle_{\mu' V T }^{r}$ calculates the average number of linkers in the system $V$ containing two fixed mDNACCs with the center-to-center distance $r$ and the chemical potential $\mu'$. At $\mu \to -\infty$, there is no linker in the system, and $F(-\infty,r)$ does not depend on $r$. Therefore, the effective interaction between two mDNACCs is
\begin{equation}\label{eff_num}
\begin{aligned}
U_{\mathrm{eff}}(r) & = F(\mu,r) - F(\mu,\infty) \\
& = \int_{-\infty}^{\mu} \left( \langle N_l \rangle_{\mu' V T }^{\infty} - \langle N_l \rangle_{\mu' V T }^{r} \right) d\mu'.
\end{aligned}
\end{equation}
 As our theory is applicable at $R \gg l \gg r_c$, we simulate the system with $l=0.2R$ and $r_c=0.01R$.

To evaluate Eq.~\ref{eff_num}, we perform grand canonical ($\mu VT$) MC simulations, in which we fix the chemical potential of linkers and position of two particles in the system, and the number of DNA linkers changes under the control of chemical potential. Besides the conventional grand canonical MC moves employed to add/remove linkers in the system, we also perform MC trial moves to randomly translate and rotate the free linkers.
For bonded linkers with one free end, we perform two kinds of trial moves to speed up the equilibration: randomly rotate around the bonded end or rotate around the center of the bonded particle.
For the binding/unbinding moves we use the approach described in Ref.~\cite{martinez2010anomalous,martinez2011designing}. For these MC trial moves, we first randomly select a linker end in the adsorption layer, if the end is in the unbound state, a binding trial move is proposed; if the end is in the bound state an unbinding trial move is proposed.
The formation of a new linker end-ssDNA bond is accepted with a probability
\begin{equation}
P_{acc}=\min{\left[1, \frac{n_S}{\lambda+1} \exp{(-\beta\Delta G_{bind})} \right]},
\end{equation}
where $n_S$ is the total number of unbound ssDNA receptors on the surface of the particle , $\lambda$ is the total number of bonds formed. Conversely, the acceptance probability of an unbinding trial move is:
 \begin{equation}
P_{acc}=\min{\left[1, \frac{1}{(n_S+1)(n_L+1)} \exp{(\beta\Delta G_{bind})} \right]},
\end{equation}
where $n_L$ is the total number of linker ends in the unbound state in the adsorption layer.

\section{Supplementary Materials}
\noindent
Supplementary material for this article is available at XXXXX-XXXXX
\\
\begin{small}
\noindent
Section S1. Derivation of combinational term $W(\{m_i,q_{ij}\})$ (Eq.~2).\\

\noindent
Section S2. Partition functions for linkers in different states.\\

\noindent
Section S3. Derivation of self-consistent equation and proof of positive definiteness.\\

\noindent
Section S4. Mean field theory for linker-mediated mDNACCs at the strong binding limit $\beta \Delta G_{bind} \to -\infty$.
\end{small}


\begin{thebibliography}{37}%

\bibitem{Jacobs:2016kg}
	W. M. Jacobs and D. Frenkel, ``Self-Assembly of Structures with Addressable Complexity,'' J. Am. Chem. Soc. \textbf{138}, 2457–2467 (2016).
	
\bibitem{alivisatos1996organization}
A. P. Alivisatos, K. P. Johnsson, X. Peng, T.E. Wilson, C. J. Loweth, M. P. Bruchez Jr, and P. G. Schultz, ``Organization of `nanocrystal molecules’ using DNA,'' Nature \textbf{382}, 609 (1996).

\bibitem{mirkin1996dna}
C. A. Mirkin, R. L. Letsinger, R. C. Mucic, and J. J. Storhoff, ``A DNA-based method for rationally assembling nanoparticles into macroscopic materials,'' Nature \textbf{382}, 607 (1996).

\bibitem{seemanjacs1996}
X. Li, X. Yang, J. Qi, and N. C. Seeman, ``Antiparallel DNA double crossover molecules as components for nanoconstruction,'' J. Am. Chem. Soc. \textbf{118}, 6131 (1996).

\bibitem{dnareviewscience}
M. R. Jones, N. C. Seeman, and C. A. Mirkin, ``Programmable materials and the nature of the DNA bond,'' Science \textbf{347}, 1260901 (2015).

\bibitem{dnabrickscience}
Y. Ke, L. L. Ong, W. M. Shih, and P. Yin, ``Threedimensional structures self-assembled from DNA bricks,'' Science \textbf{338}, 1177 (2012).

\bibitem{yinpeng2017}
L. L. Ong, N. Hanikel, O. K. Yaghi, C. Grun, M. T. Strauss, P. Bron, J. Lai-Kee-Him, F. Schueder, B. Wang, P. Wang, J. Y. Kishi, C. A. Myhrvold, A. Zhu, R. Jungmann, G. Bellot, Y. Ke, and P. Yin, ``Programmable self-assembly of three-dimensional nanostructures from 10,000 unique components,'' Nature \textbf{552}, 72 (2017).

\bibitem{Jacobs:2015gi}
W. M. Jacobs, A. Reinhardt, and D. Frenkel, ``Rational design of self-assembly pathways for complex multicomponent structures,'' Proc. Natl. Acad. Sci. USA \textbf{112}, 6313–6318 (2015).

\bibitem{Sajfutdinow2018}
M. Sajfutdinow, W. M. Jacobs, A. Reinhardt, C. Schneider, and D. M. Smith, ``Direct observation and rational design of nucleation behavior in addressable self-assembly,'' Proc. Natl Acad. Sci. USA \textbf{115}, E5877 (2018).

\bibitem{crocker2006}
A. J. Kim, P. L. Biancaniello, and J. C. Crocker, ``Engineering DNA-mediated colloidal crystallization,'' Langmuir \textbf{22}, 1991 (2006).

\bibitem{gang2008}
D. Nykypanchuk, M. M. Maye, D. van der Lelie, and O. Gang, ``Dna-guided crystallization of colloidal nanoparticles,'' Nature \textbf{451}, 549 (2008).

\bibitem{mirkin2011}
R. J. MacFarlane, B. Lee, M. R. Jones, N. Harris, G.C. Schatz, and C. A. Mirkin, ``Nanoparticle superlattice engineering with DNA,'' Science \textbf{334}, 204 (2011).

\bibitem{crocker2012}
M. T. Casey, R. T. Scarlett, W. B. Rogers, I. Jenkins, T. Sinno, and J. C. Crocker, ``Driving diffusionless trans- formations in colloidal crystals using DNA handshaking,'' Nat. Commun. \textbf{3}, 1209 (2012).

\bibitem{wang2015}
Y. Wang, Y. Wang, X. Zheng, E. Ducrot, M.-G. Lee, G.-R. Yi, M. Weck, and D. J. Pine, ``Synthetic strategies toward DNA-coated colloids that crystallize,'' J. Am. Chem. Soc. \textbf{137}, 10760 (2015).

\bibitem{wang2015natcomm}
Y. Wang, Y. Wang, X. Zheng, E. Ducrot, J. S. Yodh, M. Weck, and D. J. Pine, ``Crystallization of DNA-coated colloids,'' Nat. Commun. \textbf{6}, 7253 (2015).

\bibitem{gang2015}
Y. Zhang, S. Pal, B. Srinivasan, T. Vo, S. Kumar, and O. Gang, ``Selective transformations between nanoparticle superlattices via the reprogramming of DNA-mediated interactions,'' Nat. Mater. \textbf{14}, 840 (2015).

\bibitem{crocker2017}
Y. Wang, I. C. Jenkins, J. T. McGinley, T. Sinno, and J. C. Crocker, ``Colloidal crystals with diamond symmetry at optical lengthscales,'' Nat. Commun. \textbf{8}, 14173 (2017).

\bibitem{zeravcic2014size}
Z. Zeravcic, V. N. Manoharan, and M. P. Brenner, ``Size limits of self-assembled colloidal structures made using specific interactions,'' Proc. Natl. Acad. Sci. USA \textbf{111}, 15918–15923 (2014).

\bibitem{vanderMeulen:2013dc}
S. A. J. van der Meulen and M. E Leunissen, ``Solid Colloids with Surface-Mobile DNA Linkers,'' J. Am. Chem. Soc. \textbf{135}, 15129–15134 (2013).

\bibitem{vanderMeulen:2015jb}
S. A. J. van der Meulen, G. Helms, and M. Dogterom, ``Solid colloids with surface-mobile linkers,'' J. Phys. Condens. Matter \textbf{27}, 233101–15 (2015).

\bibitem{Hu:2018ev}
H. Hu, P. Sampedro Ruiz, and R. Ni, ``Entropy Stabilizes Floppy Crystals of Mobile DNA-Coated Colloids,'' Phys. Rev. Lett. \textbf{120}, 048003 (2018).

\bibitem{AngiolettiUberti:2014kl}
S. Angioletti-Uberti, P. Varilly, B. M. Mognetti, and D. Frenkel, ``Mobile Linkers on DNA-Coated Colloids: Valency without Patches,'' Phys. Rev. Lett. \textbf{113}, 128303– 5 (2014).

\bibitem{angioletti2013communication}
S. Angioletti-Uberti, P. Varilly, B. M. Mognetti, A. V. Tkachenko, and D. Frenkel, ``Communication: A simple analytical formula for the free energy of ligandreceptor-mediated interactions,'' J. Chem. Phys. \textbf{138}, 021102 (2013).

\bibitem{Varilly:2012gl}
P. Varilly, S. Angioletti-Uberti, B. M. Mognetti, and D. Frenkel, ``A general theory of DNA-mediated and other valence-limited colloidal interactions,'' J. Chem. Phys. \textbf{137}, 094108–16 (2012).

\bibitem{eoshd}
A. Santos, M. L{\'{o}}pez de Haro, and S. Bravo Yuste, ``An accurate and simple equation of state for hard disks,'' J. Chem. Phys. \textbf{103}, 4622 (1995).

\bibitem{depletion_AO}
S. Asakura and F. Oosawa, ``Interaction between particles suspended in solutions of macromolecules,'' J. Polym. Sci. \textbf{33}, 183 (1958).

\bibitem{depletionbook}
H. N. W. Lekkerkerker and R. Tuinier, \emph{Colloids and the Depletion Interaction} (Springer Netherlands, 2011).

\bibitem{kinetic2016a}
L. Parolini, J.Kotar, L. Di-Michele, and B.M. Mognetti, ``Controlling self-assembly kinetics of DNA-functionalized liposomes using toehold exchange mechanism,'' ACS Nano \textbf{10}, 2392 (2016).

\bibitem{kinetic2016b}
S. J. Bachmann, J. Kotar, L. Parolini, A. Saric, P. Cicuta, L. Di Michele, and B. M. Mognetti, ``Melting transition in lipid vesicles functionalised by mobile DNA linkers,'' Soft Matter \textbf{12}, 7804 (2016).

\bibitem{titojcp}
N. B. Tito, S. Angioletti-Uberti, and D. Frenkel, ``Communication: Simple approach for calculating the binding free energy of a multivalent particle,'' J. Chem. Phys. \textbf{144}, 161101 (2016).

\bibitem{lowensohn2019linker}
J. Lowensohn, B. Oyarz{\'u}n, G. N. Paliza, B. M. Mognetti, and W. B. Rogers, ``Linker-mediated phase behavior of DNA-coated colloids,'' Phys. Rev. X \textbf{9}, 041054 (2019).

\bibitem{Derjaguin1934}
B. Derjaguin, ``Untersuchungen{\"u} berdiereibungund adh{\"a}sion, iv,'' Kolloid-Zeitschrift 69, 155–164 (1934).

\bibitem{AngiolettiUberti:2012gi}
S. Angioletti-Uberti, B. M. Mognetti, and D. Frenkel, ``Re-entrant melting as a design principle for DNA-coated colloids,'' Nat. Mater. \textbf{11}, 518–522 (2012).

\bibitem{rogers2015programming}
W. B. Rogers and V. N. Manoharan, ``Programming colloidal phase transitions with DNA strand displacement,'' Science \textbf{347}, 639–642 (2015).

\bibitem{chaikin2012}
K.-T. Wu, L. Feng, R. Sha, R. Dreyfus, A. Y. Grosberg, N. C. Seeman, and P. M. Chaikin, ``Polygamous particles,'' Proc. Natl Acad. Sci. USA \textbf{109}, 18731 (2012).

\bibitem{martinez2010anomalous}
F. J. Martinez-Veracoechea, B. Bozorgui, and D. Frenkel, ``Anomalous phase behavior of liquid–vapor phase transition in binary mixtures of DNA-coated particles,'' Soft Matter \textbf{6}, 6136–6145 (2010).

\bibitem{martinez2011designing}
F. J. Martinez-Veracoechea and D. Frenkel, ``Designing super selectivity in multivalent nano-particle binding,'' Proc. Natl Acad. Sci. USA \textbf{108}, 10963–10968 (2011).

\end{thebibliography}

%

\ \\
\begin{acknowledgements}
\textbf{Acknowledgements:}
We thank Dr. Bortolo Mognetti, Dr. Qunli Lei, and Prof. Yufeng Wang for helpful discussions. {We also thank two anonymous referees for their constructive suggestions.}
This work has been supported in part by the Singapore Ministry of Education through the Academic Research Fund MOE2017-T2-1-066 (S) and (M4011873.120), by Nanyang Technological University Start-Up Grant (NTU-SUG: M4081781.120), by the Advanced Manufacturing and Engineering Young Individual Research Grant (A1784C0018) and by the Science and Engineering Research Council of Agency for Science, Technology and Research Singapore. We thank NSCC for granting computational resources.
\textbf{Author contribution:}
R.N. conceived and directed the research; X.X. formulated the mean field theory and performed numerical simulations with help from H.H.; and all authors discussed the results and wrote the manuscript.
\textbf{Competing interests:} The authors declare that they have no competing interests. \textbf{Data and materials availability:} All data needed to evaluate the conclusions in the paper are present in the paper and/or the Supplementary Materials. Additional data related to this paper may be requested from the authors.
\end{acknowledgements}

\end{document}